\documentclass[twocolumn,showpacs,jap,preprintnumbers,superscriptaddress,amsmath,amssymb]{revtex4}
\usepackage{graphicx}
\usepackage{dcolumn}
\usepackage{bm}
\usepackage{color}

\begin{document}

\preprint{Journal of Applied Physics}

\title{Cryogenic instrumentation for fast current measurement in a silicon single electron transistor}

\author{T. Ferrus}\affiliation{Hitachi Cambridge Laboratory, J. J. Thomson Avenue, CB3 0HE, Cambridge, United Kingdom}
\email{taf25@cam.ac.uk}
\author{D. G. Hasko}\affiliation{Cavendish Laboratory, University of Cambridge, J. J. Thomson Avenue, CB3 0HE, Cambridge, United Kingdom}
\author{Q. R. Morrissey}\affiliation{Rutherford Appleton Laboratory, Chilton, Didcot, OX11 0QX, Oxon, United Kingdom}
\author{S. R. Burge}\affiliation{Rutherford Appleton Laboratory, Chilton, Didcot, OX11 0QX, Oxon, United Kingdom}
\author{E. J. Freeman}\affiliation{Rutherford Appleton Laboratory, Chilton, Didcot, OX11 0QX, Oxon, United Kingdom}
\author{M. J. French}\affiliation{Rutherford Appleton Laboratory, Chilton, Didcot, OX11 0QX, Oxon, United Kingdom}
\author{A. Lam}\affiliation{Cavendish Laboratory, University of Cambridge, J. J. Thomson Avenue, CB3 0HE, Cambridge, United Kingdom}
\author{L. Creswell}\affiliation{Cavendish Laboratory, University of Cambridge, J. J. Thomson Avenue, CB3 0HE, Cambridge, United Kingdom}
\author{R. J. Collier}\affiliation{Cavendish Laboratory, University of Cambridge, J. J. Thomson Avenue, CB3 0HE, Cambridge, United Kingdom}
\author{D. A. Williams}\affiliation{Hitachi Cambridge Laboratory, J. J. Thomson Avenue, CB3 0HE, Cambridge, United Kingdom}
\author{G. A. D. Briggs}\affiliation{Department of Materials, University of Oxford, Parks Road, Oxford OX1 3PH, United Kingdom}

\date{\today}     

\begin{abstract}

We present a realisation of high bandwidth instrumentation at cryogenic temperatures and for dilution refrigerator operation that possesses advantages over methods using radio-frequency single electron transistor or transimpedance amplifiers. The ability for the low temperature electronics to carry out faster measurements than with room temperature electronics is investigated by the use of a phosphorous-doped single-electron transistor. A single-shot technique is successfully implemented and used to observe the real time decay of a quantum state. A discussion on various measurement strategies is presented and the consequences on electron heating and noise are analysed.

\end{abstract}

                          
                            
\maketitle

\section{INTRODUCTION.}

The ability to carry out high speed current measurements on nanoscale solid state devices is essential for quantum information, in implementations where rapid decoherence limits the timescale for readout.\cite{Decoherence} Conventionally, such a current measurement would make use of an ideal room temperature operating transimpedance amplifier and the measurement speed would be limited by a low pass filter effect due to the output impedance of the nanodevice and the capacitance of the connecting cable. In measurements at low temperature, this cable capacitance is significant due to the large length between the room temperature transimpedance amplifier and the nanodevice. This combined with the high output impedance of semiconductor single electron transistors \cite{SET} results in a practical measurement bandwidth of a few 100\,Hz. Alternatively, a high speed conductance measurement may be carried out using radiofrequency techniques.\cite{RFSET} This offers a much higher measurement bandwidth, but requires the nanodevice to be connected to the room temperature measurement system using high bandwidth cables. In previous work,\cite{Hasko} it has been shown that a low pass filter is required on all connections to the nanodevice in order to prevent electron heating. The electron heating is manifest in degradation of the Coulomb oscillations. Another measurement technique is to integrate the current onto a capacitor for a fixed period of time and then to measure the increase in voltage across the capacitor. This approach has much greater complexity compared to the transimpedance amplifier, since the capacitor must be periodically discharged. But, such a function may be implemented without the use of high value resistors and so can operate under cryogenic conditions. By operating the measurement circuit at low temperature, the cable to the nanodevice can be considerably reduced in length, so that the low pass filter effect is diminished and a much higher measurement bandwidth is possible. In this paper, we describe the use of a custom complementary metal-oxide-semiconductor integrated circuit that operates at 4.2\,K (LTCMOS) and is compatible with dilution refrigerator operation. The capabilities of this measurement circuit are investigated through the characterisation of a phosphorous-doped silicon single electron transistor (SET) \cite{Fulton,Beenakker} which is known to be sensitive to microwave irradiation.\cite{Creswell} A significant improvement in measurement speed was found compared to the conventional approach using room temperature transimpedance amplifiers ($\sim10^4$ times faster \cite{Gorman}). Finally we show that such an approach enables the real time evolution of system states to be monitored by a single shot technique \cite{Single shot} due to the reduction in microwave heating compared to the conventional approach.

\section{Low temperature instrumentation and test device}
\subsection{Fast measurement techniques}

A transimpedance amplifier outputs a voltage proportional to its input current and can easily be built from an operational amplifier and a single resistor. However, in a non-ideal transimpedance amplifier, the measurement bandwidth is limited by the time constant resulting from the feedback resistor of the amplifier and the stray capacitance. The value of this resistor increases with the sensitivity so that the amplifier generally imposes a measurement speed limit for the lower current ranges. Operating the transimpedance amplifier at low temperature reduces the capacitance but causes a considerable variation in the characteristics of the high value resistors used in the feedback circuit and a significant heat loading especially when in close proximity to the device.\cite{HEMT}

Alternatively, a radiofrequency (RF)-based SET gives a measurement bandwidth up to few tens of MHz. The SET device is embedded into a resonant circuit and the RF reflectivity from the source contact is measured. The variation of charge in the device then modifies the conductance of the SET so modulating the amplitude of the reflected wave. Because of the high frequency operation, the $1/f$ noise, due to the background charge, is reduced.  However, these methods cannot be used with devices such as phosphorous-doped silicon SETs that are sensitive to microwaves.\cite{Hasko, Creswell} Indeed the use of RF induces a significant increase in the electron temperature of the device but filtering is not applicable in this situation. Also RF may induce irreversible charge movements or charge tunnelling in the SET island that affect capacitances. 

Small currents may also be measured using a charge integration circuit. By avoiding the need for a high value resistance, the charge integrator is made more compatible with cryogenic operation. The operation of a charge integrator is significantly more complicated than that of a transimpedance amplifier. Switching circuits are needed to periodically discharge the integration capacitor and these additional circuits are conveniently accommodated within a purpose made application specific device, designated for cryogenic operation. By appropriate choice of the integration capacitance and switching times, the charge integration circuit may be operated at significantly greater speed than an equivalent transimpedance amplifier.

\subsection{Measurement setup}

\subsubsection{LTCMOS ASIC}

The measurement unit consists of a custom LTCMOS Application Specific Integrated Circuit (ASIC) and a room temperature Data AcQuisition (DAQ) and communication box. The LTCMOS ASIC supplies the gate ($V_{\textup{g}}$) and source-drain voltages ($V_{\textup{SD}}$) and measures the SET drain current ($I_{\textup{SD}}$). A high speed and sensitive current measurement is achieved by using the SET current to charge a very small value capacitor for a predetermined time. The voltage level on the capacitor is then sampled by the DAQ and communication box, and then discharged periodically. 100\,fF, 1\,pF, and 10\,pF capacitors and sampling periods as low as 25\,$\mu$s may be selected under computer control to adjust the speed and sensitivity of the measurement. The maximum sampling time and input current are limited by the saturation of the capacitor voltage measurement circuit. The measurement resolution and noise are a function of signal level, sampling time, and measuring circuit capacitance. Parasitic input capacitances (such as that due to the cable between the LTCMOS ASIC and the device being tested) also degrade the noise performance. For improved measurement quality it is therefore desirable to minimise the cable length to the SET and use a cryogenic measurement system. The DAQ and communications box at room temperature acts as a controller for the LTCMOS ASIC plus a data digitiser and recorder for the integrated signal data that is sent back from the LTCMOS ASIC. This data is then transmitted via gigabit optical Ethernet to a host computer for storage and analysis (Fig. 1). The measurement units are integrated into a custom designed probe that is constructed from thin walled stainless steel tubing to minimise heat transfer from the room temperature parts into the cryostat. The probe is intended to be inserted into a dilution unit and is divided into a number of levels corresponding to the structures of an Oxford Instruments Superconductivity KelvinOx$^{\textup{TM}}$ 400 dilution refrigerator. Copper plates and clamps at the mixing chamber, cold plate, still, 1\,K pot and 4\,K plate levels allow thermal coupling of signal cables to avoid thermal loading to lower levels. The heat load from the LTCMOS ASIC is also transferred to the helium bath via the 4.2\,K plate by this method. A custom designed device carrier is mounted at the bottom of the probe to hold the device package and microwave feed and provide thermal coupling to the mixing chamber.

\subsubsection{Mounting for microwave experiments}

Devices were mounted onto 20 pin leadless header packages that are suitable for signal frequencies up to $\sim$1\,GHz, according to the manufacturer specifications. However the inductance due to the bond wires (30\,$\mu$m diameter and 1\,mm length) and the resistance of the doped silicon leads (sheet resistance $\sim$1\,k$\Omega$.$\Box^{-1}$), in combination with the stray capacitance to the substrate, causes a significant attenuation at high frequencies. While this bandwidth is more than sufficient to carry the gate, source and drain voltages as well as current measurement signals, it is not fast enough to carry a microwave manipulation pulse, which may include signal frequencies between 1 to 20\,GHz so that the microwave coupling to the device needs to be implemented differently. In previous work, this difficulty has been overcome by embedding the device in a microwave printed circuit board and by using coplanar waveguides for the high frequency signals.\cite{coplanar} However, it is difficult to match the impedance to the one of the device in this approach, due to the problem of providing a 50\,$\Omega$ terminating resistance close to the nanostructure. As a result, standing wave effects make the microwave power coupled to the device depend on frequency. In addition, it is known that the signal leads need to be low-pass filtered (typically preventing frequencies above $\sim$1\,MHz from propagating to the device) to avoid significant degradation of the Coulomb characteristics through electron heating (see Sec. II. C. 2.). The approach used here employs an open ended semi-rigid cylindrical waveguide (RG402) to weakly couple microwaves in the frequency range from $\sim$1-20\,GHz to the device in a global manner. By controlling the distance between the device and the end of the waveguide standing wave effects may be avoided. The base of the header package is a thick gold layer (much thicker than the skin depth in the relevant frequency range) and so provides an efficient reflector of microwaves. Part of the incident wave is reflected back from the base of the header package into the waveguide with a phase change of $\pi$ radians from the incoming signal. The remainder of the microwave signal is able to pass between the end of the waveguide and the header package base. This part of the microwave signal is also subject to a reflection due to the propagation from an environment with 50\,$\Omega$ waveguide impedance to an environment with a characteristic impedance of 377\,$\Omega$.$\Box^{-1}$. This second reflection is in phase with the incoming signal in contrast to the reflection from the base of the header package. By adjusting the gap between the end of the waveguide and the base of the header package ($\sim$1\,mm), these two reflections can be matched in amplitude so that the cavity behaves as a 50\,$\Omega$ termination. As this effect is determined by geometry rather than electrical properties, the impedance matching is independent of frequency and temperature over a wide range.

\subsubsection{Microwave Experimental setup}

Measurements were controlled by a LabView program running on a computer that integrated the LTCMOS interface, with an Agilent E8257D-520 PSG Analog Signal Generator and an Agilent B1130A pulse pattern generator linked via a GPIB bus (Fig. 1). The microwave source provides continuous wave (cw) signals to 1\,Hz precision at power levels of up to +15\,dBm; the frequency may be changed on a time-scale of a few milliseconds. The cw microwave signal is passed through a Herotek single pole single throw S1D0518A4 PIN switch, which has a switching time of about 30\,ns. The PIN switch is controlled by a B1130A Pulse Pattern Generator (PPG), allowing arbitrary on-off switching patterns to be implemented within the limitations imposed by the switching time of the PIN switch and the memory capacity of the PPG. A trigger signal generated by the DAQ and communication box is used to synchronize the start of a sequence of current measurements with the initiation of a pulse sequence from the PPG. Most of the signal path makes use of flexible SMA terminated microwave cable so that some signal loss occurs. The microwave power available at the device from this setup was sufficient for cw measurements, but for pulsed measurements an additional amplification stage, based on a Mini-Circuits ZX60-5916M-S preamplifier and an output stage using a BFM21 FET, was used to increase the microwave power to about +25\,dBm (in the range up to a few GHz). Most of the measurements were performed at 4.2\,K by immersing the probe in a liquid helium dewar for preliminary assessment and to provide an efficient thermal coupling to the 4.2\,K bath. Complementary experiments were carried in the internal vacuum chamber of the dilution refrigerator but placed additional constraints on power dissipation especially during microwave experiments.

\begin{figure}
\begin{center}
\includegraphics[width=86mm]{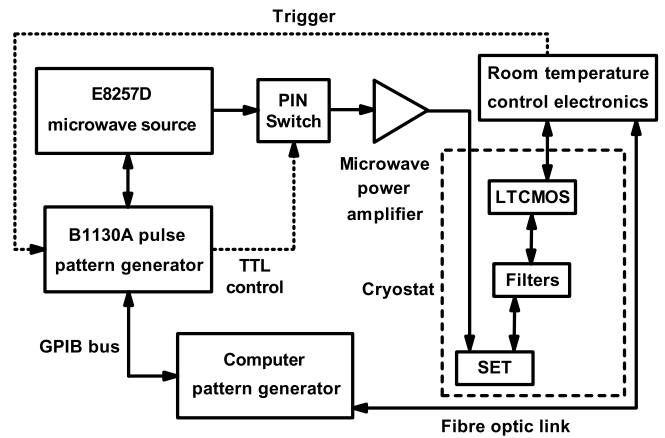}
\end{center}
\caption{\label{fig:figure1} Schematic of the measurement setup. Arrows indicates directional communication between apparatus.}
\end{figure}

\subsection{Device fabrication and DC characteristics}
\subsubsection {SET as a test device}

A highly doped silicon SET was chosen to test the LTCMOS since the characteristics of these devices have been widely explored previously using a variety of room temperature measurement systems. The SETs used in this work are fabricated from a standard silicon-on-insulator (SOI) substrate having a 30 nm-thick highly-doped silicon layer. Phosphorous atoms were ion implanted, resulting in a post-annealing concentration of $\sim\,10^{20}$\,cm$^{-3}$. High resolution electron beam lithography and reactive ion etching were used to pattern a single island of diameter $\sim$60\,nm. The 30-nm width constrictions that separate the island to the connecting source and drain leads are depleted of electrons and act as tunnel barriers. The formation of such barriers has been widely investigated and will not be discussed here.\cite{Vitkavage} Thermal oxidation is then used to reduce the island size and tunnel barrier widths as well as for surface passivation, resulting in a significant reduction in random telegraph switching. The devices are controlled by an in-plane gate that is formed using the same SOI layer. Several devices were processed identically and behaved similarly. The following results are presented for a specific device that was chosen for its high reproducibility in time as well as for its high signal to noise ratio.

\subsubsection {DC characteristics and electrical noise heating}

All measurements were performed using the 1\,pF current measurement capacitor in the LTCMOS ASIC which gave the best signal-to-noise ratio (SNR) of the three possible values. The visibility of Coulomb blockade oscillations at 4.2\,K is poor with a SNR of about 18:1 for $V_{\textup{SD}}$=2\,mV and a significant background conductivity (Fig. 2) when compared with conventional room temperature equipment. From the dependencies in gate and source drain bias, it is possible to extract estimates for the gate and drain level arms, e.g. $\alpha_{\textup{g}}$ and $\alpha_{\textup{D}}$:

\begin{eqnarray}\label{eqn:equation1}
\alpha_{\textup{g}}=C_{\textup{g}}/C_{\Sigma} \mbox{ and } \alpha_{\textup{D}}=C_{\textup{D}}/C_{\Sigma}
\end{eqnarray}

where $C_{\textup{g}}$, $C_{\textup{D}}$ and $C_{\Sigma}$ are the gate, drain and total capacitance extracted from the Coulomb diamond slopes.

We find $\alpha_{\textup{D}} \sim 0.53$ and $\alpha_{\textup{g}} \sim 8.7 \times 10^{-3}$ and that both values are approximately independent of gate voltage. The SET diameter determined from the scanning electron microscope (SEM) is about 60\,nm, so that the first excited state in the dot is expected to be at 1.1\,meV above the ground state, a value somewhat larger than the thermal broadening at helium temperature ($\sim 0.4$\,meV). Under this condition, the electron temperature $T_{\textup{e}}$ can be estimated from :

\begin{eqnarray}\label{eqn:equation2}
\alpha_{\textup{g}}\sim 3.5 k_{\textup{B}}T_{\textup{e}}/\left(e W\right) 
\end{eqnarray}

where $W$ is the width of the Coulomb peak.

This approximation is valid as long as the intrinsic linewidth of the Coulomb peak is negligible compared with the thermal broadening, e.g. $T>4.2\,$K in the present case. For $V_{\textup{g}}$=-0.039\,V and $V_{\textup{SD}}$=2\,mV, we find $T_{\textup{e}} \sim 9.0$\,K implying significant electron heating. 

In the absence of deliberate microwave irradiation, high frequency electromagnetic noise was found to be the major source of electron heating in the present experimental setup. Apart from electrical noise coming from the LTCMOS itself, the room temperature control electronics that contains digital circuitry operating at clock speeds of 25\,MHz and 125\,MHz (plus a 1\,GHz internal clock to the optical transceiver) may contribute to the noise that propagates down to the device. For equivalent measurements, made using conventional room temperature electronics, such as Keithley 236 source-measurement units for example, this problem may be circumvent by using 2\,MHz cut-off frequency low pass filters (MiniCircuits BLP1.9) on each lead connecting the device. Unfortunately, the high capacitance of the BLP1.9 circuit prevents its use with the charge integrating circuit of the LTCMOS circuit as the noise gain is greatly increased. In this case, we used a single stage resistor-capacitor low-pass filter made of a ceramic capacitor and a metal film resistor because of the materials being temperature independent. To extend the high frequency cut-off, ferrite beads were attached to the filter inputs and outputs. The ferrite material was chosen so as to suppress noise over a wide range of frequencies as well as to enhance its efficiency at low temperature. Filters were placed near the device to remove electromagnetic noise as well as to thermally anchor the leads. Unlike a normal inductor, ferrite beads absorb the energy and dissipate it as heat. However the amount of heat is negligible and does not contribute to electron heating effects. The cut-off frequency at 4.2\,K was 2\,MHz so that single shot measurements were not affected by the filters. The use of this cryogenic filter improved the SNR of the Coulomb blockade oscillations to 64:1 and resulted in the appearance of blockade regions with zero current.

For $V_{\textup{g}}$=-0.083\,V and $V_{\textup{SD}}$=2\,mV, we find $\alpha_{\textup{D}} \sim 0.44$, $\alpha_{\textup{g}} \sim 3.9 \times 10^{-2}$ $T_{\textup{e}} \sim 4.6$\,K and the charging energy was about 4.3\,meV. We also find $C_{\textup{g}}\sim 2.0$\,aF and $C_{\Sigma}\sim 37.0$\,aF and a corresponding dot diameter of about 57\,nm, assuming a spherical dot. This value is in good agreement with the dot size determined from SEM imaging. The electron temperature was thus efficiently reduced to a value close to the lattice temperature during liquid helium immersion measurement (Fig. 2).

For measurements in the dilution refrigerator, the sample is located in a internal vacuum chamber so that the lower operating temperature reduces the thermal conductivity and specific heat capacity in the device. These effects make the problem of electron heating in the dilution refrigerator much more severe than for liquid helium immersion measurements. A base temperature, measured by the mixing chamber thermometer, of 
27\,mK was obtained for dc measurements and 60\,mK for microwave experiments due the additional heat loading from the waveguide. The electron temperature was estimated to be $T_{\textup{e}}\sim300\,$mK for $V_{\textup{SD}}$=-2.25\,mV, 150\,mK for -1.75\,mV and about 80\,mK for 1.25\,mV after correction for the source drain offset of 0.75\,mV for dc measurements.

\begin{figure}
\begin{center}
\includegraphics[width=86mm]{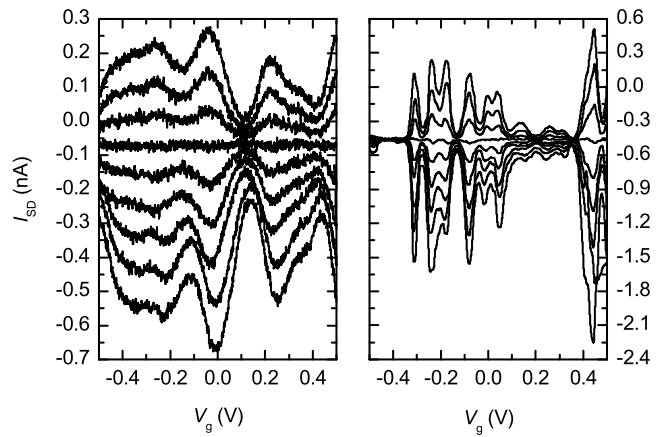}
\end{center}
\caption{\label{fig:figure2} Coulomb blockade oscillations when lines are not filtered (left) and after adding the low temperature $RLC$ filter to the circuit (right). The measurement was taken at 4.2\,K with source-drain biases from -2\,mV (bottom) to 2\,mV (top).}
\end{figure}

\subsubsection {Continuous microwave spectroscopy and microwave heating}

In previous work, the CW microwave excitation has been found to consist of a large number of resonances that are detectable with the LTCMOS when the microwave power $P$ at the source is greater than about -20\,dBm (see Fig. 3a and b).\cite{Hasko} These resonances exhibit a wide range of amplitudes and quality factors with the largest amplitudes corresponding to the lowest $Q$ values (Fig. 3c). Far from being a unique case, the presence of resonances with high $Q$ value have also been observed in TiO$_2$ sol-gel based high-k gate dielectric silicon MOSFETs where they result from resonant excitation between trapped states.\cite{Khan} Previously resonant features have also been observed in a SET. Manscher noticed an increase of source-drain current in an Al/Al$_2$O$_3$/Al SET as a result of enhancement of the cotunneling effect due to microwave induced heating.\cite{Manscher} However, in the resonances observed here, we noticed the presence of both a relative increase and a relative decrease in the SET current, which is inconsistent with an explanation based on heating. Also, in our setup, a variation in the microwave power does not induce a relative change in the current through a modification of the microwave coupling constant because of the method used to couple the microwave to the device. Finally, figure 3c shows that a same mechanism is responsible for both high and low $Q$ resonances.In the case of a highly doped silicon SET, the resonances are likely to result from a tunnelling process that is well isolated from the environment (Fig. 3b). From the Lorentzian shape of the resonances and their width that is five or six orders of magnitude lower than $k_{\textup{B}}T$, it has been shown that the most likely electronic process that is taking place is phononless transport like in frequency driven two-level tunneling systems \cite{Shklovskii} or trap assisted tunnelling via localised states in or around the SET island.

These localised states are favoured by the presence of phosphorous donor in the device despite their metallic-like density. Because of surface segregation of dopants from the silicon that was consumed by thermal oxidation, a significant number of donors are embedded into a silicon oxide matrix leading to numerous electrons being localised in states on the sidewalls of the SET.\cite{defect} This effect is enhanced by the large surface state density found on the non-(100) surfaces forming the sidewalls. The presence of these trapped electrons induces sidewall depletion that may result in the creation of tunnel barriers in small width silicon region. The electron displacement, due to a tunnelling process occurring between localized states, modifies the electrostatic environment in the SET island so that the magnitude of the measured current is changed. Localised states in the direct transport pathway are expected to lead to low $Q$ resonances with large amplitude due to the short residence time of the electron. Whereas more remote states are expected to lead to the higher $Q$, but lower amplitude, resonances. Electron displacement can lead to an increase or decrease (or combination of these) in the SET current depending on the proximity and orientation of the localized states with the respective tunnel barriers to the island. Coupling to the phonon bath is weak and electronic processes happen over a timescale shorter than the thermalization time, so that the resonance widths $\Delta\nu_{\textup{0}}$ could be related to the energy loss mechanism in the system using $Q=\pi \nu_{\textup{0}}/\Delta\nu_{\textup{0}}\,\sim\,\pi T_{\textup{0}}\nu_{\textup{0}}$, where $T_{\textup{0}}$ is the energy relaxation time of the system \cite{lifetime} and $\nu_{\textup{0}}$ the resonance frequency.

In the Coulomb blockade regime, the conductivity oscillations are known to be sensitive to electron temperature through a change in the Coulomb peak lineshape and an increase in the background conductivity. As microwave irradiation inevitably leads to electron heating, the visibility of the oscillations can be used to determine the maximum usable microwave power before noticeable heating occurs (Fig. 4a). As demonstrated by Manscher,\cite{Manscher} in a Al/A$_2$O$_3$/Al SET, the co-tunneling effect is modified in the presence of an ac signal and the conductivity is linear (respectively quadratic) with microwave power in the conducting (respectively blockade) regime to the second order perturbation (Fig. 4b and c).
Away from the resonance conditions as described previously, most of the microwave absorption occurs in the substrate by coupling to the low electron density in the lightly doped silicon. Because of the efficient coupling between the silicon substrate and the phonon bath, the temperature of the silicon will then rise until the thermal losses match the incoming microwave power.

We experimentally found that a $10\,\%$ increase in the temperature of the silicon substrate would lead to a noticeable change in the Coulomb blockade oscillations. In the case of a direct immersion into liquid helium at 4.2\,K, the thermal impedance is dominated by the Kapitza resistance between the semiconductor and the liquid cryogen.\cite{Kapitza} From the value of the Kapitza coefficient between silicon and $^4$He \cite{Olson} and for a device dimension of 2.5 mm $\times$ 2.5 mm $\times$ 0.25 mm, we estimate that power needed to cause such an increase of temperature is about 4\,mW for that geometry. As the cooling power of helium gas is much greater than for the liquid, this may underestimate the heat load required since even a modest rise in temperature would result in boiling off liquid helium. The situation in the dilution refrigerator is very different. As the sample is in vacuum, the heat load is taken mainly by the sample wires (200\,$\mu$m diameter and 10\,cm long), which are thermally anchored to the mixing chamber. At the lowest operating temperature (about 60\,mK) and by using the value of thermal conductivity for pure copper, the minimum power is reduced down to 0.1\,$\mu$W. The lower thermal conductivity to the thermal anchoring and the reduced tolerance to temperature rise at the lower temperatures in the dilution then greatly reduce the maximum power that can be coupled to the device.

\begin{figure}
\begin{center}
\includegraphics[width=85mm]{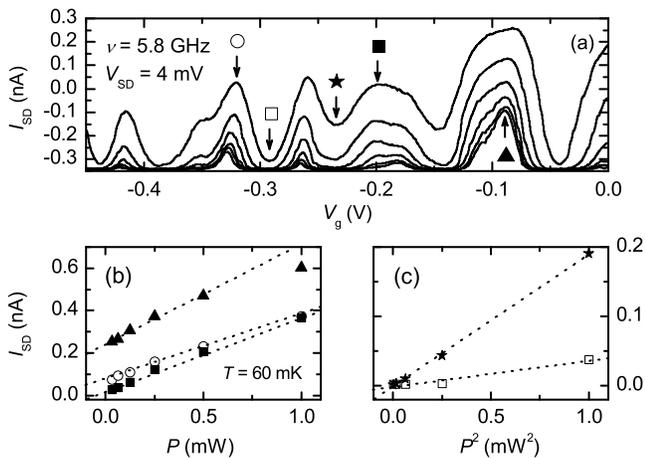}
\end{center}
\caption{\label{fig:figure4} a) Gate oscillations at 4.2\,K under microwave illumination with $P=0$\,dBm (top) down to -15\,dBm (bottom). b) Variation of the current with microwave power in the conducting region( Coulomb blockade maxima as designed by symbols in a)). c) Variation in the blockade regime where cotunneling is expected to be the dominant transport mechanism. Symbols refer to maxima and minima in the Coulomb blockade oscillations as designed in a).}
\end{figure}

\section{An example of fast measurement : the single-shot technique.}
\subsection{Single shot measurement}

The high $Q$ value and resonant frequency suggests that the lifetime for the associated excitation can be as long as 100$\,\mu$s for some resonances. With the larger measurement bandwidth afforded by the use of the LTCMOS, the real-time evolution of the source-drain current after the application of a short microwave pulse can be measured. This technique is also known as a single shot measurement. This procedure consists of sending a single  microwave excitation pulse of duration $t_p$ whilst carrying out current measurements at intervals $t_m$ with the condition $t_p\leq  t_m\ll \tau\ll t_r$ where $t_r$ the time between repeated single shot operations and $\tau$ the quantum state lifetime. To this purpose, a well-isolated resonance on a level background at a frequency $\nu_{\textup{0}}\sim 2.718$\,GHz was selected from a cw measurement at a power of +9\,dBm, with a charge integration time of 200$\,\mu$s and averaging the results over 200 measurements (Fig. 5). 

This results in a signal-to-noise ratio of about 100:1. The width of the chosen resonance was about 207\,kHz, for which the corresponding relaxation time is $T_{\textup{0}}\sim 5\,\mu$s. A single shot measurement was then performed by reducing the charge integration time to 20\,$\mu$s and using a microwave pulse of duration of 2\,$\mu$s. The microwave power was increased to about +25\,dBm using a power amplifier to compensate for the decrease in the signal amplitude. These conditions also greatly reduced the SNR compared to the cw measurement, so we proceeded to a to a numerical average over 1000 identical and independent single-shot measurements. The delay in the trigger signal causes the microwave pulse to arrive at $t\sim 0.8\,$ms (Fig. 5). The signal measured in single shot mode also contains some very large damped oscillations, which do not depend on the nature of the microwave pulse. They are even present in the absence of the microwave pulse and are probably due to broadband radiofrequency excitation due to the digital signals in the control circuits. By tuning the microwave frequency to on or off resonance, the effects of these background oscillations may be removed. The difference between these is an exponential decay with a time constant $\tau\sim 90\,\mu$s. This value is more than one order of magnitude longer than that found from the cw linewidth measurement. As the time-averaged power in the pulsed measurement is a factor of 1000 less than that used in the cw measurement, this difference in relaxation time may result from the much greater microwave-induced electron heating in the cw measurement. The lifetime $\tau$ corresponds to the time for the microwave-excited electron to relax to its original state or to escape. Long lifetimes are consistent with glassy behaviour,\cite{glass} a characteristic manifestation of Coulomb interaction in systems with localized states. The disorder in the highly doped silicon system is probably due to the random placement of phosphorous donors. The dynamics of transport and relaxation in this system results from the competition between disorder and Coulomb interactions. 

\begin{figure}
\begin{center}
\includegraphics[width=85mm]{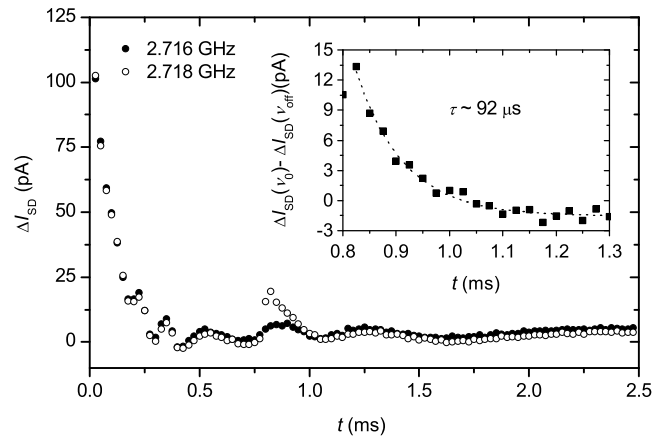}
\end{center}
\caption{\label{fig:figure5} Time evolution of the relative SET current off resonance ($\nu_{\textup{off}}=2.716\,$GHz) and on resonance ($\nu_{\textup{0}}=2.718\,$GHz). The inset shows the real-time decay of the current.}
\end{figure}

\subsection{Measurement methods comparison}
\subsubsection{Single shot spectroscopy versus cw and pulse microwaves}

A comparison of the lifetime $T_{\textup{0}}$ deduced from the resonance linewidth in a cw measurement and the lifetime $\tau$ obtained from a single shot measurement, indicates that the $Q$ value may be considerably underestimated by cw spectroscopy. Indeed, the continuous application of microwaves, at power levels needed for signal detection, significantly raises the electron temperature and so reduces the visibility of the resonances due to the enhancement of thermally driven relaxation and consequently the $Q$ value. This problem is more acute in a dilution refrigerator compared with the liquid helium immersion experiment because of the difference in specific heat capacity and thermal conductivity of the system as discussed in section II. C. 3. The strategy to maintain a reasonable SNR during single shot measurements is quite complex and generally requires the time delay to be adjusted between to successive measurements. This reduces the overall measurement speed but allow the device to exchange heat with the mixing chamber through the copper wires. In single shot measurements, the time averaged microwave power is greatly decreased by using pulses, even though the instantaneous power is significantly higher than in the cw case. This problem is circumvented by using the single-shot approach for spectroscopy. The signal amplitude at each frequency, is taken from the first current measurement point after the arrival of the microwave pulse, in single shot mode, i.e. at $t=825\,\mu$s. This method resulted in the observation of resonances with $Q$ factors as high as 400,000 measured in the dilution refrigerator and at a temperature of 60\,mK (Fig.6) with negligible microwave induced heating. 

\begin{figure}
\begin{center}
\includegraphics[width=85mm]{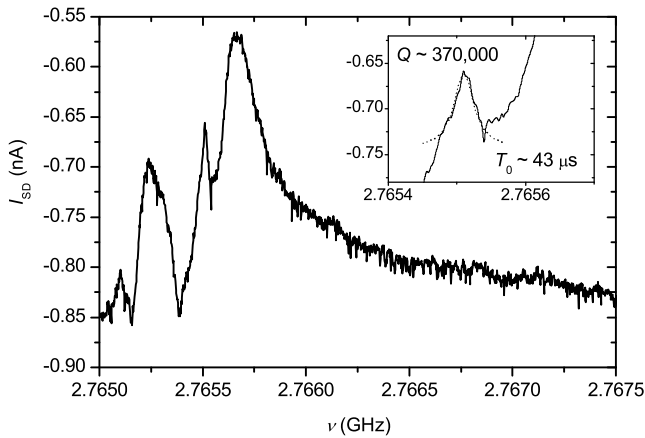}
\end{center}
\caption{\label{fig:figure6} Single-shot mode spectroscopy showing a very high $Q$ resonance at 60\,mK. Measurements were taken with $P=$7\,dBm, a gate voltage of 34\,mV, a source voltage of 2\,mV and a pulse duration of 2\,$\mu$s}
\end{figure}

\subsubsection{Single shot technique versus standard time dependence measurement}

Other methods are proposed in order to implement time dependence measurements and may have even higher bandwidth. This is the case of RFSETs \cite{RFSET} where the time dependence is extracted from current spectroscopy by the use of a Fourier transform. This method allows ultra-fast measurement with a bandwidth of the order of 100\,MHz, but cannot be used in devices that are sensitive to radio-frequency signals as is the case for doped or undoped silicon SETs.\cite{PSi} Also, high frequencies are incompatible with the signal filtering that is essential in reducing the electron temperature.

A time average measurement has also been used when the evolution of a quantum state could not be monitored in real-time due to measurement bandwidth limitations. In quantum computing architectures such as Gorman's, \cite{Gorman} this method requires the measured current to be averaged over a large number of identical gate voltage pulses, with the condition that the next pulse is applied before the excitation, due to the previous pulse, can fully decay.\cite{pulses} This limits the time between successive pulses and assumes long term stability of the system under investigation. By contrast, the single shot measurement technique allows the real-time evolution of the system to be recorded after excitation by a single pulse, starting and finishing with the system in a fully relaxed state.

\section{CONCLUSIONS.}

We have realised a high bandwidth low temperature CMOS amplifier circuit based on charge integration for the use in dilution refrigerators. This allowed to implement the single shot measurement technique and observe slow-decaying electronic states in a phosphorous-doped silicon single electron transistor. In order to address the issue of electron heating usually occurring during continuous microwave spectroscopy, we have developed a single shot spectroscopy technique that permits the observation of high quality factor resonances up to 400,000 at 60\,mK that were not observable by other methods.

\section{Acknowledgements.}

The authors thank EPSRC (UK) for funding under grant number GR/S24275/01 and for a Professorial Research Fellowship for G. A. D. B. GR/S15808/01. They are also grateful to V. Mikheev and P. Noonan from Oxford Instrument for their involvement in the design and the realisation of the dilution refrigerator. This work was partly supported by Special Coordination Funds for Promoting Science and Technology in Japan. Two of the authors (T. F. and D. G. H.) contributed equally to this work.


\end{document}